\documentclass[12pt]{article}
\begin{document}

\vspace{-35mm}

\title{LABORATORY MODELLING OF THE EARTH RADIATION BELT}

\author{
{Dr. Alexander LUCHINSKIY}$^{a}$ \\[0.5cm]
{Prof. Dr. Yakov S. SHIFRIN}$^b$ \\[3em]
  {\normalsize (a) Johannes Gutenberg - University Mainz, }\\
  {\normalsize     Institut f\"ur Physik, Germany }\\
  {\normalsize     Gymnasialstr 11, D-55543 Bad Kreuznach, Germany }\\
  {\normalsize     E-mail: Aluchinsk@aol.com} \\
  {\normalsize     Tel./Fax  +49 671 / 35594} \\[.5em]
  {\normalsize (b) Kharkov State Technical University of Radio Electronics }\\
  {\normalsize             Lenin Avenue 14, Kharkov, 310726 Ukraine }\\
  {\normalsize     E-mail: shifrin@kture.kharkov.ua} \\
  {\normalsize     Tel. +380 572 409430 } \\
  {\normalsize     Fax  +380 572 409113 } 
}
\date{}
\maketitle

\begin{abstract}
 Method of the laboratory modelling of the Earth radiation belt is
 presented. Method can be used for the estimation of consequences
 of global energetic and communication projects realizations.
\end{abstract}

\vspace*{10pt}

The radiation belt of the Earth is the inner part of the magnetosphere, in which the 
geomagnetic field hold charged particles with kinetic energy from 10 KeV to 100 MeV.
This belt play the essential role for the sun radiation regime for the Earth and for 
the electromagnetic waves propagation.

As it is well known, even small alteration of solar activity influent essential upon 
the biological and ecological balance at the Earth for all levels - from viruses and 
micro-organisms to biological societies and ecological systems at all. The most bright 
indicator of the solar activity alteration are changes of the pathogenic organisms activity 
as a result of a displacement of biological equilibrium. But the other phenomena, which are 
not connected with pathogenic biological objects are very important for the life on the 
Earth and ecological balance too, although they can not be so obviously observed.

In this century the realisation of some global projects will be very actual for the supply 
of the Earth with the electrical energy to be converted from the solar energy and for the 
creation of global communication systems. The realisation of these projects cannot be neutral 
for the ionosphere and for the radiation belt state.

For example, for the Glazer Project it will be necessary to create a system of powerful 
solar energy converters, which should send powerful SHF beams through the radiation belt. 
But even small alterations in the radiation belt will change the solar irradiation regime 
of the Earth the same way, as for alteration of solar activity by constant magnetosphere.

Thus it is very dangerous to realise global energetic and communication projects without 
the all-round investigation of consequences of these realisations.

That is why it is necessary in particular to have a possibility of the radiation belt 
laboratory modelling and its experimental investigation as a complete system under the 
influence of different factors.

Besides that this task is actual also immediate for the solving of problems of radiation 
propagation and communication too.

This problem was solved in \cite{luch4} and \cite{luch5} for a general case, i.e. for all 
kinds of planets, which have magnetical  fields.  The laboratory Modelling of  the Earth 
radiation belt is in fact a particular case of this solution.  Therefore no additional 
patent descriptions for any details will be required.   

In the Earth radiation belt the charged particles realised a motion, which can be considered 
as a sum of three components: 1 - quick cycle motion due to Lorenz force in the geomagnetic 
field; 2 - the "longitude-drift" of these cycles along the Equator in the crossed magnetic 
and  gravitational fields; and 3 - the oscillation of particles in directions, normal to 
the cycles planes and to the direction of drift.

Change of movement directions in these oscillations occurs due to geometrical phenomenon 
of so-called specular reflection.

The main problem for the realisation of radiation belt laboratory modelling is that it is 
on principle impossible to create the 3-dimensional radial gravitation field in a 
laboratory device. 

This problem can be solved through the replacing of gravitation field with the electrostatical 
field, because the mathematical description of these two fields are similar. For the realisation 
of the drift-phenomenon it is principally necessary the existence of two crossed fields. 
It is not important for this process, what kind of field will be used together with the 
magnetic field, gravitational field or the electrostatical one.    

Finally it is proposed to use for this modelling the Penning Trap \cite{luch1} or our 
physically similar traps with other construction of electrodes \cite{luch2}-\cite{luch5}, 
which can make this modelling more effective.

By doing this way, the cycle motion in Earth magnetic field corresponds to the "cyclotron 
motion" in the Penning Trap, the drift along the Equator corresponds to the "magnetron motion", 
and the interpolar oscillation corresponds to the "axial motion".          

For the realisation of laboratory experiments with this model it is necessary to provide the 
possibility of the ceaseless change of field configuration during the experiment. It is achieved 
through the special construction of the trap ring electrode. This electrode in the form of 
rotation-hyperboloid of one sheet is collected from straight plates, which lies in hyperbolic 
surface. This technical solution can be used not only for the aims of the problem under 
consideration, but in constructions of some antenna systems with smooth rearrangeable 
geometry too.


\begin{thebibliography}{99}
\bibitem{luch1}
Geonium theory: physics of a single electron or ion in a Penning trap;
L. S. Brown, G. Gabrielse; Reviews of Modern Physics, vol. 58, No 1, 1986.

\bibitem{luch2} 
DE 198 26 241 A1 Int. Cl.6 H 01 J 49/42 Eine Einrichtung zur dynamischen 
Speicherung von Teilchen; Dr. A. Luchinskiy, Prof. Dr. G. Werth, A. Drakoudis; 1998

\bibitem{luch3} DE 198 25 604 A1 Int. Cl.6 H 01 J 49/42; G 01 N 27/00 
Eine Einrichtung zur Realisierung speziell geformter Laserfelder in einer 
Quadrupol Teilchenfalle; Dr. A. Luchinskiy, Prof. Dr. G. Werth, N. Hermanspahn; 1998

\bibitem{luch4} 
DE 100 09 238 A 1 Int. Cl.7 G 09 B 23/06  Verfahren der Labormodellierung 
des Strahlungsg\"urtels der Planeten; Dr. A. Luchinskiy; 2000.

\bibitem{luch5} 
Deutsche Patentanmeldung Nr. 101 53 235.0 Verfahren der Labormodellierung des 
Strahlungsg\"urtels der Planeten; Dr. A. Luchinskiy; 2001.
\end{thebibliography}
\end{document}